\documentclass[preprint]{aastex}
\usepackage[colorlinks,citecolor=blue,linkcolor=blue,urlcolor=blue,dvips]{hyperref}
\usepackage{lineno}
\usepackage{ulem}
\nolinenumbers

\newcommand{\be}{\begin{equation}}
\newcommand{\ee}{\end{equation}}
\newcommand{\bea}{\begin{eqnarray}}
\newcommand{\eea}{\end{eqnarray}}

\author{
G.~M.~Madejski\altaffilmark{1}, 
K.~Nalewajko\altaffilmark{2,1}, 
K.~K.~Madsen\altaffilmark{3},
J.~Chiang\altaffilmark{1}, 
M.~Balokovi\'{c}\altaffilmark{3}, 
D.~Paneque\altaffilmark{19}, 
A.~K.~Furniss\altaffilmark{1,4},
M.~Hayashida\altaffilmark{5}, 
C.~M.~Urry\altaffilmark{6}, 
M.~Sikora\altaffilmark{2},
M.~Ajello\altaffilmark{7}, 
R.~D.~Blandford\altaffilmark{1}, 
F.~A.~Harrison\altaffilmark{3}, 
D. Sanchez\altaffilmark{8},
B. Giebels\altaffilmark{9},
D.~Stern\altaffilmark{10}, 
D.~M.~Alexander\altaffilmark{11},
D.~Barret\altaffilmark{12,13},
S.~E.~Boggs\altaffilmark{14},
F.~E.~Christensen\altaffilmark{15},
W.~W.~Craig\altaffilmark{14,16},
K.~Forster\altaffilmark{3},
P.~Giommi\altaffilmark{17},
B.~Grefenstette\altaffilmark{3},
C.~Hailey\altaffilmark{18},
A.~Hornstrup\altaffilmark{15},
T.~Kitaguchi\altaffilmark{20},
J.~E. Koglin\altaffilmark{1},
P.~H.~Mao\altaffilmark{3},
H.~Miyasaka\altaffilmark{3},
K.~Mori\altaffilmark{18},
M.~Perri\altaffilmark{17,21},
M.~J.~Pivovaroff\altaffilmark{16},
S.~Puccetti\altaffilmark{17,21},
V.~Rana\altaffilmark{3},
N.~J.~Westergaard\altaffilmark{15},
W.~W.~Zhang\altaffilmark{22},
A.~Zoglauer\altaffilmark{14}
}

\altaffiltext{1}{Kavli Institute for Particle Astrophysics and Cosmology, Department of Physics and SLAC National Accelerator Laboratory, Stanford University, Stanford, CA 94305, USA}
\altaffiltext{2}{Nicolaus Copernicus Astronomical Center, Polish Academy of Sciences, Bartycka 18, 00-716 Warsaw, Poland}
\altaffiltext{3}{Cahill Center for Astronomy and Astrophysics, Caltech, Pasadena, CA 91125, USA}
\altaffiltext{4}{California State University - East Bay, 25800 Carlos Bee Boulevard, Hayward, CA 94542}
\altaffiltext{5}{Institute for Cosmic Ray Research, University of Tokyo, 5-1-5 Kashiwanoha, Kashiwa, Chiba, 277-8582, Japan}
\altaffiltext{6}{Yale Center for Astronomy and Astrophysics, Physics Department, Yale University, PO Box 208120, New Haven, CT 06520-8120, USA}
\altaffiltext{7}{Department of Physics and Astronomy, Clemson University, Kinard Lab of Physics, Clemson, SC 29634-0978, USA}
\altaffiltext{8}{Laboratoire d`Annecy-le-Vieux de Physique des Particules, Universite de Savoie, CNRS/IN2P3, F-74941 Annecy-le-Vieux, France}
\altaffiltext{9}{Laboratoire Leprince-Ringuet, Ecole Polytechnique, CNRS/IN2P3, F-91128 Palaiseau, France}
\altaffiltext{10}{Jet Propulsion Laboratory, California Institute of Technology, Pasadena, CA 91109, USA}
\altaffiltext{11}{Department of Physics, Durham University, Durham DH1 3LE, UK}
\altaffiltext{12}{Universite de Toulouse, UPS - OMP, IRAP, Toulouse, France}
\altaffiltext{13}{CNRS, Institut de Recherche en Astrophysique et Planétologie, 9 Av. Colonel Roche, BP 44346, F-31028 Toulouse Cedex 4, France}
\altaffiltext{14}{Space Science Laboratory, University of California, Berkeley, CA 94720, USA}
\altaffiltext{15}{DTU Space, National Space Institute, Technical University of Denmark, Elektrovej 327, DK - 2800 Lyngby, Denmark}
\altaffiltext{16}{Lawrence Livermore National Laboratory, Livermore, CA 94550, USA}
\altaffiltext{17}{ASI Science Data Center, Via del Politecnico snc I - 00133, Roma, Italy}
\altaffiltext{18}{Columbia Astrophysics Laboratory, Columbia University, New York, NY 10027, USA}
\altaffiltext{19}{Max-Planck-Institut fur Physik, D-80805 Munchen, Germany}
\altaffiltext{20}{RIKEN, Nishina Center, 2 - 1 Hirosawa, Wako, Saitama, 351 - 0198, Japan}
\altaffiltext{21}{INAF - Osservatorio Astronomico di Roma, via di Frascati 33, I - 00040 Monteporzio, Italy}
\altaffiltext{22}{NASA Goddard Space Flight Center, Greenbelt, MD 20771, USA}

\title{First {\it NuSTAR} observations of the BL Lac - type blazar PKS~2155-304:  constraints on the jet content and distribution of radiating particles}

\begin{abstract}
We report the first hard X-ray observations with {\it NuSTAR} of the BL Lac type blazar PKS~2155-304, 
augmented with soft X-ray data from XMM-{\it Newton} and $\gamma$-ray data from the {\it Fermi} Large Area Telescope, 
obtained in April 2013 when the source was in a very low flux state.  
A joint {\it NuSTAR} and XMM spectrum, covering the energy range 0.5 - 60 keV, 
is best described by a model consisting of a log-parabola component 
with curvature $\beta = 0.3^{+0.2}_{-0.1}$ and a (local) photon index $3.04\pm 0.15$ at photon energy 
of $2\;{\rm keV}$, and 
a hard power-law tail with photon index $2.2\pm 0.4$. The hard X-ray tail can be smoothly 
joined to the quasi-simultaneous $\gamma$-ray spectrum by a synchrotron self-Compton  
component produced by an electron distribution with index $p = 2.2$.
Assuming that
the power-law electron distribution extends down to $\gamma_{\rm min} = 1$
and that there is one 
proton per electron, an unrealistically high total jet power of 
$L_p \sim 10^{47}\;{\rm erg\,s^{-1}}$ is inferred.
This can be reduced by two orders of magnitude either by considering a significant presence
of electron-positron pairs with lepton-to-proton ratio 
$n_{\rm e+e-}/n_{\rm p} \sim 30$, or by introducing an additional, 
low-energy break in the electron energy distribution at the electron Lorentz factor $\gamma_{\rm br1} \sim 100$.  
In either case, the jet composition is expected to be strongly matter-dominated.
\end{abstract}

\begin{document}

\section{Introduction}

PKS~2155-304 is one of the most extensively studied BL Lac objects.  It is a strong emitter of 
electromagnetic radiation in all observable bands, from radio to very high energy (VHE) 
$\gamma$ rays.  Its $E \times F(E)$ broad-band spectrum reveals two prominent peaks 
located respectively in the far UV/soft X-ray band, and in the multi-GeV part of the high 
energy $\gamma$-ray band.  As such, PKS~2155-304 belongs to the class of jet-dominated active 
galaxies with the jet pointing close to our line of sight - known as blazars - 
and, specifically, to a sub-class known as high-energy peaked BL Lac 
objects, or HBLs (see, e.g., \citealt{Pad95}).  

The two-peak spectral energy distribution (SED) of HBL blazars is generally (and most successfully) 
interpreted in the context of leptonic synchrotron self-Compton (SSC) models \citep[e.g.,][]{Ghi98}, 
where the low-energy component is presumably due to synchrotron emission, while the high energy 
component is due to inverse Compton scattering by the same electrons that produce the synchrotron 
peak.  The optical spectra of the HBL blazars are generally devoid of emission lines even 
in the low jet flux states, implying a rather weak isotropic radiation field associated 
with the accretion.  In such objects, it is generally believed that the dominant population 
of ``seed'' photons (as seen in the co-moving frame of the relativistic jet) are the 
synchrotron photons produced within the jet.

From an observational standpoint, in HBL-type blazars perhaps the least is known about 
the lowest-energy part of the inverse-Compton peak.  This is primarily due to the limited sensitivity 
of instruments in the relevant energy range, from $\sim 20$ keV to $\sim 100$ MeV.  
In particular, the onset of the high-energy peak contains important information about the 
lowest-energy electrons in the jet, which, in the context of any emission model, are most 
plentiful, and thus are a sensitive probe of the total content of particles 
in the jet.  Notably, this low-energy end of the electron population cannot be reliably 
studied in the synchrotron component, since at low energies, the synchrotron emission is 
likely self-absorbed.  Fortunately, the successful launch of the {\it NuSTAR} mission, sensitive in the 
3 - 79 keV energy range, opened a new window for sensitive searches for the low-energy ``tail'' of 
the electron distribution in the inverse Compton component.  

In this paper, we report {\it NuSTAR} observations of PKS~2155-304, one of the brightest 
and also most luminous HBL blazars.  This object, at $z=0.116$, has been known 
as a bright X-ray emitter since its discovery by HEAO-1 A3 \citep{Sch79}.
Subsequent X-ray observations consistently show soft X-ray spectra, 
with photon index $\Gamma > 2.5$  in the 2 - 10 keV band 
\citep[e.g.,][]{Sem93,Bri94,Ede95,Urr97,Zha99,Kat00,Tan01,Bhag14}.
Rapid variability on hourly time scales in the X-ray and optical bands 
is common;  see \citep{Zha99,Ede01,Tan01,Kat00}.  PKS~2155-304 is a known bright VHE 
$\gamma$-ray source \citep{Cha99,Aha05} and is highly variable on timescales 
down to $\sim$ minutes in the VHE $\gamma$ rays \citep{Aha07}.  
For the most recent multi-band observations involving {\it Fermi}-LAT and VHE 
observatories, see \cite{Aha09} or \cite{Che15}.

{\it NuSTAR} observed PKS~2155-304 multiple times in 2013, as a part of multi-frequency 
monitoring with ground-based observatories, spanning radio through VHE bands.  
Here, we focus on the X-ray spectroscopy afforded by the first observation, conducted 
strictly simultaneously with XMM-{\it Newton}, in April 2013 for cross-calibration purposes.  
The joint {\it NuSTAR} and XMM-{\it Newton} spectrum reveals spectral 
complexity, and specifically, a soft spectrum in the 2 -- 10 keV range, hardening 
at the high-energy part of its bandpass.  While a similar hard spectral ``tail'' 
was previously measured in the spectrum of this object by HEAO-1 \citep{Urr82} as well as
by {\it Beppo}-SAX \citep{Gio98}, this was done with less sensitive, non-imaging instruments;  
the sensitive {\it NuSTAR} observation allows us to reliably confirm 
its presence, and characterize the spectrum in more 
detail.  With relatively simple modeling of the broad-band SED in the context of 
SSC models, we are able to draw inferences about the distribution of radiating particles 
over a broad range of energies.  

Unless otherwise specified, we adopt the concordance cosmology, 
$\Omega_{\rm M} = 0.3, \Omega_{\Lambda} = 0.7$, and $H_{\rm 0} = 70$ km\ s$^{-1}$\ Mpc$^{-1}$.

\section{Observations and data reduction}

While {\it NuSTAR} observed PKS~2155-304 multiple times in 2013, here 
we report on the first observation, performed on 2013 April 23-24, 
or around MJD 56405.  Those observations were coordinated to be strictly 
simultaneous with multiple X-ray instruments for the purpose of cross-calibration.  
The campaign, described in Madsen et al. (2016), yielded useful data from {\it Chandra}, 
{\it Swift}, {\it Suzaku}, {\it NuSTAR}, and XMM-{\it Newton}.  For the purpose of the current study, 
we use only the XMM-{\it Newton} soft X-ray, and the {\it NuSTAR} hard X-ray data, as those 
provided the best statistics;  we also include {\it Swift} UVOT data, to provide 
simultaneous optical / UV coverage towards constraining the emission models.   
Subsequent {\it NuSTAR} observations of PKS~2155-304 were conducted simultaneously 
with the H.E.S.S.-II Cherenkov telescope, and will be reported elsewhere; for 
a preliminary overview, see \cite{San15}.

\subsection{{\it NuSTAR}}

{\it NuSTAR}, a NASA Small Explorer satellite sensitive in the hard X-ray band, 
features two multilayer-coated telescopes, focusing the reflected X-rays 
on the pixellated CdZnTe focal plane modules, FPMA and FPMB.  The observatory provides a bandpass 
of 3 -- 79 keV with spectral resolution of $\sim 1$ keV. The field of view 
of each telescope is $\sim 13'$, and the half-power diameter of an image 
of a point source is $\sim 1'$. This allows a reliable estimate and subtraction 
of  instrumental and cosmic backgrounds, resulting in an unprecedented sensitivity 
for measuring fluxes and spectra of celestial sources. For more details, see \cite{Har13}.

\begin{figure}
\includegraphics[height=\columnwidth,angle=-90]{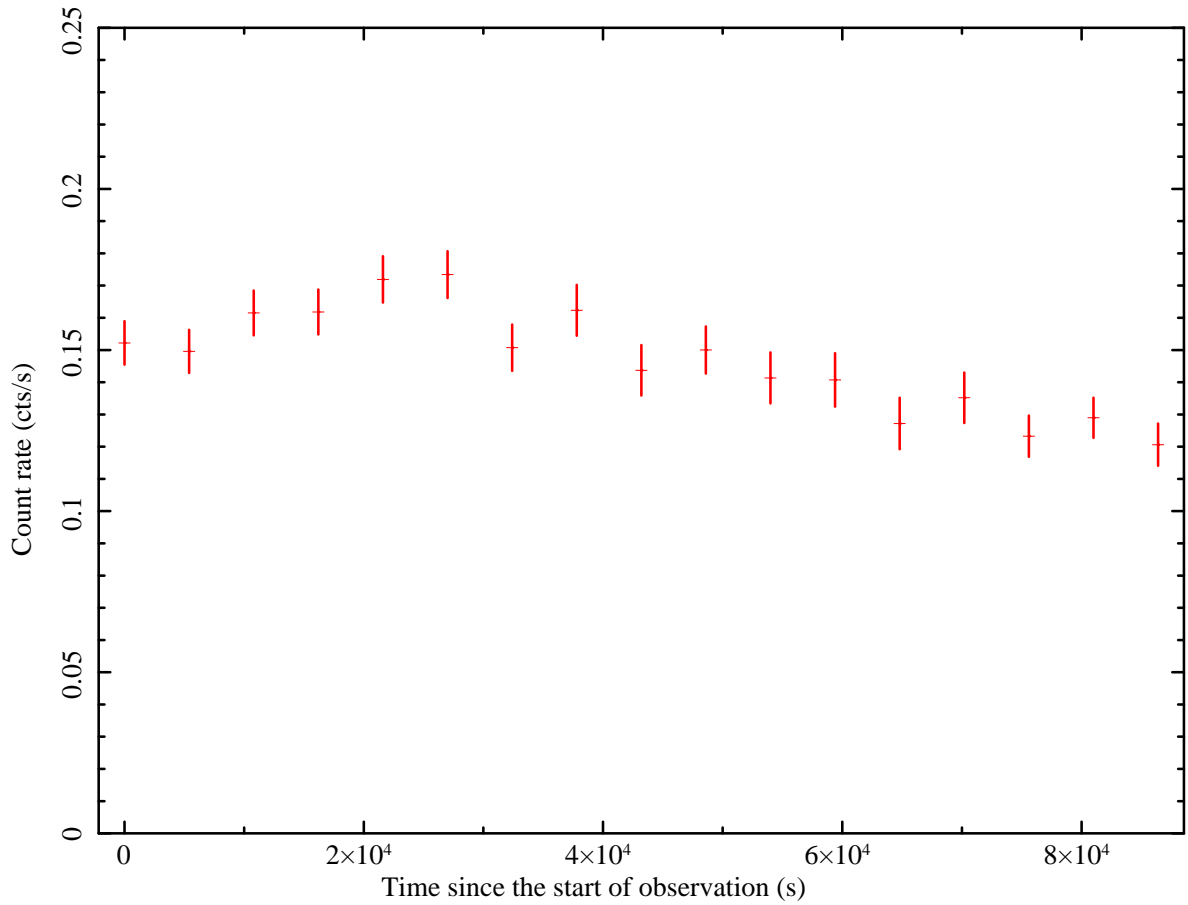}
\caption{Light curve of PKS 2155-304, as observed by {\it NuSTAR} Focal Plane module ``B" 
on 2013 April 23/24, binned on an orbital time scale.}
\label{fig1}
\end{figure}

After screening for the South Atlantic Anomaly 
passages and Earth occultation, the 2013 April 23/24 
pointing resulted in 44.9 \,ks of net observing time (OBSID 60002022002).  
The raw data products were processed with the {\it NuSTAR} Data Analysis Software 
(NuSTARDAS) package v. 1.3.1 (via the script {\tt nupipeline}), producing 
calibrated and cleaned event files.  Source data were extracted 
from a region of $45''$ radius, centered on the centroid of 
X-ray emission, while the background was extracted from a $1.5'$ radius 
region roughly $5'$ SW of the source location. Spectra were binned 
in order to have at least 30 counts per rebinned channel.  We considered the 
spectral channels corresponding nominally to the 3 -- 60 keV energy range, 
where the source was robustly detected.  The mean net (background-subtracted) 
count rates were $0.133 \pm 0.002$ and $0.129 \pm 0.002$ cts s$^{-1}$, respectively, 
for the modules FPMA and FPMB.  The raw (not background-subtracted) counts binned 
on an orbital time scale are plotted in Figure \ref{fig1}.  
The source was variable from one orbit to another, although with only a modest 
amplitude, not exceeding 10\%.  We find no change in the hardness ratio of the 
source as a function of time, indicating that there was no significant spectral variability 
during the observation.  Therefore, we focus here on time-averaged spectral analysis, 
summing the data into one deep spectral file.   

\subsection{XMM-{\it Newton} and {\it Swift} UVOT}

XMM-{\it Newton} consists of three X-ray telescopes.  Two of these 
focus celestial X-rays onto MOS CCD arrays, while the third uses the EPIC-pn 
camera.  XMM-{\it Newton} observations of PKS~2155-304 
were reduced using the XMM-{\it Newton} Science Analysis System (SAS) v. 14.0, 
with the calibration files of 2015 July 1.  The reduction followed 
exactly the same procedures as those described in Madsen et al. (2016).  
The spectra were extracted from a region $20''$ in radius for all three detectors, 
with events recorded in the inner $10''$ discarded to avoid pile-up effects.  The 
background was extracted from the corners of the the EPIC-pn for the {\tt pn} data, 
and from the empty sky fields of the peripheral CCD for the MOS data.  We considered 
the 0.5 - 10 keV energy range for spectral fitting.  In this spectral range, the count rate was 
$2.366 \pm 0.006$, $2.526 \pm 0.006$, and $6.926 \pm 0.012$ counts s$^{-1}$ and the net 
exposures were 
64770 s, 64770 s, and 66050 s, respectively, for the MOS1, MOS2, and {\tt pn} cameras.  

We also analyzed the {\it Swift} UVOT data from the pointings contemporaneous with the 
{\it NuSTAR} pointing, to ensure that those are consistent with our modelling.  
Specifically, we measured the following de-reddened fluxes at respective frequencies:   
$5.5 \times 10^{14}$ Hz:  $7.1 \pm 0.2 \times 10^{-11}$ erg cm$^{-2}$ s$^{-1}$, 
$6.9 \times 10^{14}$ Hz:  $7.3 \pm 0.2 \times 10^{-11}$ erg cm$^{-2}$ s$^{-1}$, 
$8.5 \times 10^{14}$ Hz:  $7.8 \pm 0.3 \times 10^{-11}$ erg cm$^{-2}$ s$^{-1}$, 
$11.4 \times 10^{14}$ Hz:  $7.5 \pm 0.2 \times 10^{-11}$ erg cm$^{-2}$ s$^{-1}$, 
$13.4 \times 10^{14}$ Hz:  $8.8 \pm 0.3 \times 10^{-11}$ erg cm$^{-2}$ s$^{-1}$, and 
$14.5 \times 10^{14}$ Hz:  $8.1 \pm 0.3 \times 10^{-11}$ erg cm$^{-2}$ s$^{-1}$.  
We include those in our modelling of the broad-band spectrum in Section 4.  


\subsection{{\it Fermi}-LAT}

The {\it Fermi} Large Area Telescope ({\it Fermi}-LAT; \citealt{Atw09}) is a 
pair-conversion $\gamma$-ray detector sensitive in the energy range 20 MeV to 
greater than 300 GeV.  We analyzed the {\it Fermi}-LAT data with the software 
package {\tt ScienceTools v10r0p5}, using the instrument response 
function {\tt P8R2\_SOURCE\_V6} (front and back),
including the Galactic diffuse emission model {\tt gll\_iem\_v06},
and the isotropic background model {\tt iso\_P8R2\_SOURCE\_V6\_v06}.  
Because during the {\it NuSTAR} observation on MJD~56405, PKS~2155-304 displayed a relatively 
low $\gamma$-ray state, we considered data collected over the 10-day period MJD~56400-56410, 
centered on the {\it NuSTAR} observation at MJD~56405.  Gamma-ray events were 
selected from a region of interest within $15^\circ$ of PKS~2155-304, and the background 
model includes all sources from the 2FGL catalog \citep{Nolan2012} 
within $15^\circ$ from PKS~2155-304, as well as the standard Galactic 
diffuse, isotropic and residual instrumental background emission 
models provided by the {\it Fermi} Science Support 
Center\footnote{\texttt{http://fermi.gsfc.nasa.gov/ssc/data/access/lat/Background Models.html}}. 
The photon indices of all background sources were fixed.  

The spectral data points were calculated by applying the unbinned maximum likelihood 
analysis in logarithmically spaced energy bins (with the width of the bins 
corresponding to the ratio of bin boundary energies of 2.512) 
with the photon index fixed in each bin to $\Gamma = 2$.
For each bin, we set the detection criterion to require that the test 
statistic, or ${\rm TS} \ge 10$ and $N_{\rm pred} \ge 3$.  
The source was detected (${\rm TS} > 10$) in all energy bins in the energy range 
between $\sim 250\;{\rm MeV}$ and $\sim 15\;{\rm GeV}$.  
For the bins where this criterion is not satisfied, we 
calculated the 95\% confidence level flux upper limits (i.e., 
flux $F$ such that $\log(\mathcal{L}(F)/\mathcal{L}_0) = 2$, where $\mathcal{L}_0$ is 
the best-fit likelihood value).  
(For a definition of ``Test Statistics,'' see \citealt{Mat96}).  


\section{Spectral fitting} 

\subsection{{\it NuSTAR}}

The spectral fitting of all X-ray data was performed using XSPEC v12.8.2, with the 
standard instrumental response matrices and effective area files derived using 
the {\tt ftool nuproducts}.  We fitted the data for both {\it NuSTAR} detectors simultaneously 
allowing an offset of the normalization factor for module FPMB with respect to module FPMA.  
Regardless of adopted models, the normalization offset was less than 3\%.  
First, we adopted a simple power-law model modified by 
the effects of the Galactic absorption, corresponding to a column of $1.42 \times 10^{20}$ cm$^{-2}$ 
(\citealt{Kal05}).  While the fit is acceptable ($\chi^2 = 304$ for 
295 Pulse Height Analysis, or PHA bins) 
and returns the power-law index of $2.73 \pm 0.04$, the residuals show that the {\it NuSTAR} 
spectrum is more concave (i. e. the spectrum gets flatter towards higher energies) 
than a simple power-law model would imply.   Note that this is in contrast to 
previously measured spectra of two other HBL-type 
blazars, Mkn 421 \citep{Bal15} and Mkn 501 \citep{Fur15}, where the $E \times F(E)$ 
spectra, where the  {\it NuSTAR} data augmented by {\it Swift}-XRT data allowing for a broad bandpass, 
appear to steepen with energy.  

\begin{figure}
\includegraphics[height=\columnwidth,angle=-90]{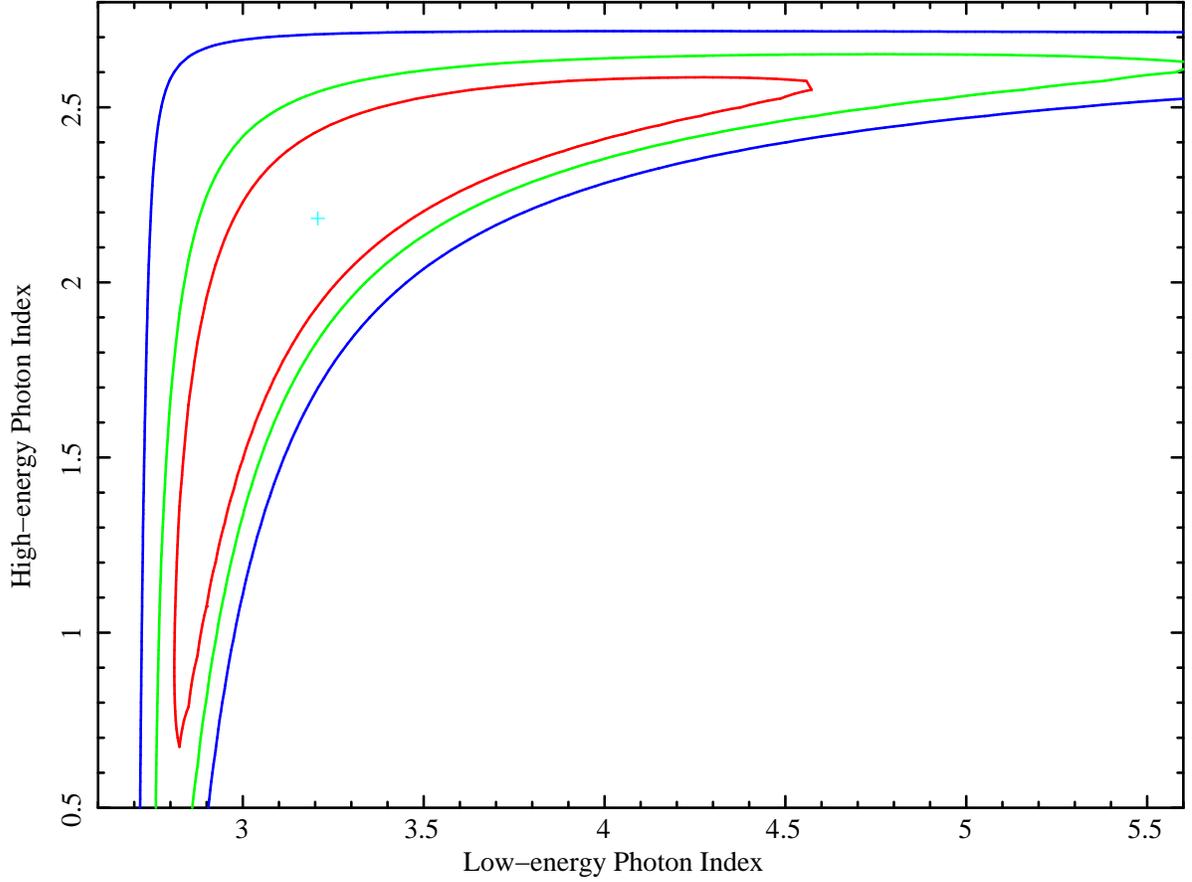}
\caption{Confidence regions for the low- vs. high-energy power-law indices determined from the {\it NuSTAR} data alone.  
The assumed model was a superposition of two power laws.  The cross corresponds to the best-fit value, while the contours from 
inside to outside correspond to the regions bounded by $\chi^{2}_{\rm min} + 2.7, 4.6$ and 9.2.  }
\label{fig2}
\end{figure}

We next attempted two more complex models (both with absorption fixed at the Galactic 
value as above).  First, we considered a broken power law, with the steeper low-energy and harder 
high-energy indices.  The fit returned significantly improved with $\chi^2 = 297$, 
or $\Delta \chi^{2}$ of 7, for 295 PHA bins.    
The low- and high-energy indices are respectively $2.82 ^{+0.12}_{-0.06} $  and $2.55 \pm 0.14$, and the 
break energy is at $8.0 ^{+2.8}_{-2.7}$ keV.  Since a broken power law model is somewhat 
unphysical, we also attempted a double power law representation of the data, also modified 
by Galactic absorption as above.  The fit returns $\chi^2 = 292$ for 295 PHA bins 
with a low-energy index of $3.03 ^{+1.1}_{-0.25}$ and a high-energy index of $1.85 \pm 0.70$.   
Given the somewhat better value of $\chi^2$, and since it can represent a superposition 
of two separate components, we express a preference for the two 
power-law model.  We plot the confidence regions of the low- versus high-energy indices for the 
two-power law model in Figure \ref{fig2}.  We also attempted to 
substitute in the place of a power law for the soft spectrum, dominating below 6 keV, 
a log-parabolic model where one additional parameter is added to allow for 
a gradual departure from a simple power law \citep[cf.][]{Tra07}.  This substitution
does not improve the quality of spectral 
fit for the {\it NuSTAR} data alone (but it {\sl does} for the joint {\it NuSTAR} + XMM-{\it Newton} 
spectral fits;  see below).  Regardless of the model, the flux of the source in 
the 2 -- 10 keV spectral band (chosen for easy comparison with previous 
observations of PKS~2155-304) is $1.1 \times 10^{-11}$ erg cm$^{-2}$ s$^{-1}$, 
which is quite faint for this source, indicating that we are observing PKS~2155-304 in a 
very low state.  For a comparison, the ``low-state'' of PKS~2155-304 
reported by \cite{Aha09} was significantly higher, ranging from $\sim 3$ to 
$\sim 9 \times 10^{-11}$ erg cm$^{-2}$ s$^{-1}$.

In order to investigate the possibility that the apparent hardening of the spectrum 
of PKS~2155-304 toward higher energies is an artifact of background subtraction, 
the analysis was repeated with multiple background regions from various regions on the detector.  
Regardless of the selected region, the departure (at high energies) from 
the very soft, $\Gamma \sim 3$ photon index persists, and we discuss the significance in 
the following Section.  

\subsection{XMM-{\it Newton} and joint {\it NuSTAR} + XMM-{\it Newton}}

We fitted all three XMM-{\it Newton} detectors simultaneously over the bandpass of 0.5 -- 10 keV.  
These data alone can be adequately fit by a model including a simple 
power law + neutral absorption:  the fit returned an equivalent hydrogen column 
of $2.6  \pm 0.2 \times 10^{20}$ cm$^{-2}$, a power-law index of $2.82 \pm 0.01$, 
and $\chi^2 = 2263$ for 2200 PHA bins.  If one imposes the fixed Galactic 
column of $1.42 \times 10^{20}$ cm$^{-2}$, the fit is significantly worse, 
with $\chi^2 = 2392$ for 2200 PHA bins.  This indicates that the source's soft X-ray spectrum 
shows significant departure from a power-law model.  Motivated by previous 
successes in applying more complex models to describe data for HBL-type blazars such 
as Mkn 421 \citep{Bal15} and Mkn 501 \citep{Fur15}, we next attempted a 
log-parabolic ({\tt logpar}) model \citep[cf.][]{Tra07}.  Such a model fits the data 
well:  $\chi^2 = 2284$ for 2200 PHA bins.  We chose the pivot energy $E_{p}$ 
to be 2 keV, close to the (logarithmic) middle of the XMM-{\it Newton} bandpass.  
The fit returns the local power-law index at that energy to be $2.80 \pm 0.02$, with 
a curvature parameter $\beta$ of $0.09 \pm 0.02$.  

The strictly simultaneous observation with XMM-{\it Newton} and {\it NuSTAR} allows an 
unprecedented (for this source) bandpass of 0.5 - 60 keV, and this is the 
bandpass we use for subsequent fits.  Motivated by the 
success of the {\tt logpar} model above, we applied it to the joint data.  
We set the absorption to the Galactic value, $1.42 \times 10^{20}$ cm$^{-2}$ as above.  With this, 
again setting the $E_{p} = 2$ keV, the best fit using all five instruments - 
XMM-{\it Newton} MOS1, MOS2, and {\tt pn} as well as {\it NuSTAR} FPMA and FPMB - returns $\chi^2 = 2629$ 
for 2495 PHA bins, with the local power-law index at 2 keV of $2.78 \pm 0.02$, 
and the curvature parameter $\beta = 0.06 \pm 0.02$.

\begin{figure}
\includegraphics[height=\columnwidth,angle=-90]{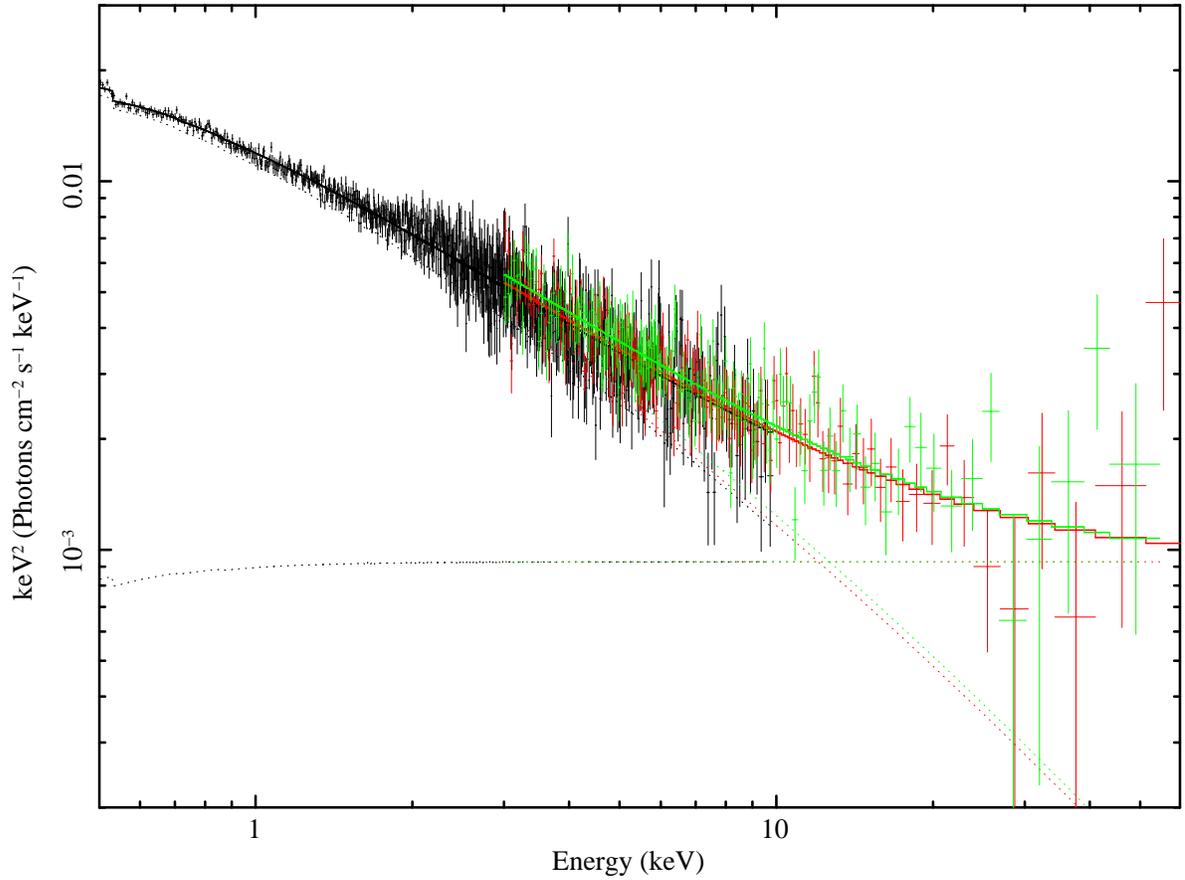}
\caption{XMM-{\it Newton} {\tt pn} and {\it NuSTAR} data for the joint observation of 
PKS~2155-304 on 2013 April 23-24 (the XMM-{\it Newton} MOS data are omitted from
the plot for clarity).  The solid line 
represents the model including the log-parabolic power 
law component, plus another, hard high-energy power law;  
the dotted lines are the two components constituting the total model.}
\label{fig3}
\end{figure}

While the above fit is acceptable, motivated by evidence for the additional hard tail 
in the {\it NuSTAR} spectrum, we attempt the final model consisting of photoelectric 
absorption by the Galactic column 
and a two-component continuum modelled as {\tt logpar} + hard power law.  The best fit 
returns $\chi^{2} = 2546$ for 2495 PHA bins.  The local index at 2 keV is now $3.04^{+0.16}_{-0.14}$, 
and $\beta = 0.3^{+0.2}_{-0.1}$.   For the high-energy (``hard tail'') power law, the index 
is $2.2 \pm 0.4$, and its normalization (at 1 keV) is $2 \times 10^{-3}$ photons 
keV${-1}$ cm$^{-2}$ s$^{-1}$ (corresponding to the 20 - 40 keV flux of $1.4 \times 10^{-12}$ erg 
cm$^{-2}$ s${-1}$).    Clearly the statistical improvement to the fit is quite pronounced, 
mainly owing to the remarkably broad bandpass provided by the combination of 
XMM-{\it Newton} and {\it NuSTAR}.  We present this final model in Fig. \ref{fig3}.  

To further test the significance of the hard tail, we performed a Monte-Carlo simulation 
of the {\it NuSTAR} + XMM data assuming just the logpar model without the power law for 1500 
realizations.  We found that none of the realizations were 
able to reproduce the feature at the observed magnitude, implying that the additional 
power-law component is significant at the 99.93\% confidence level.   


\subsection{{\it Fermi}-LAT}

\begin{figure}
\includegraphics[width=\columnwidth]{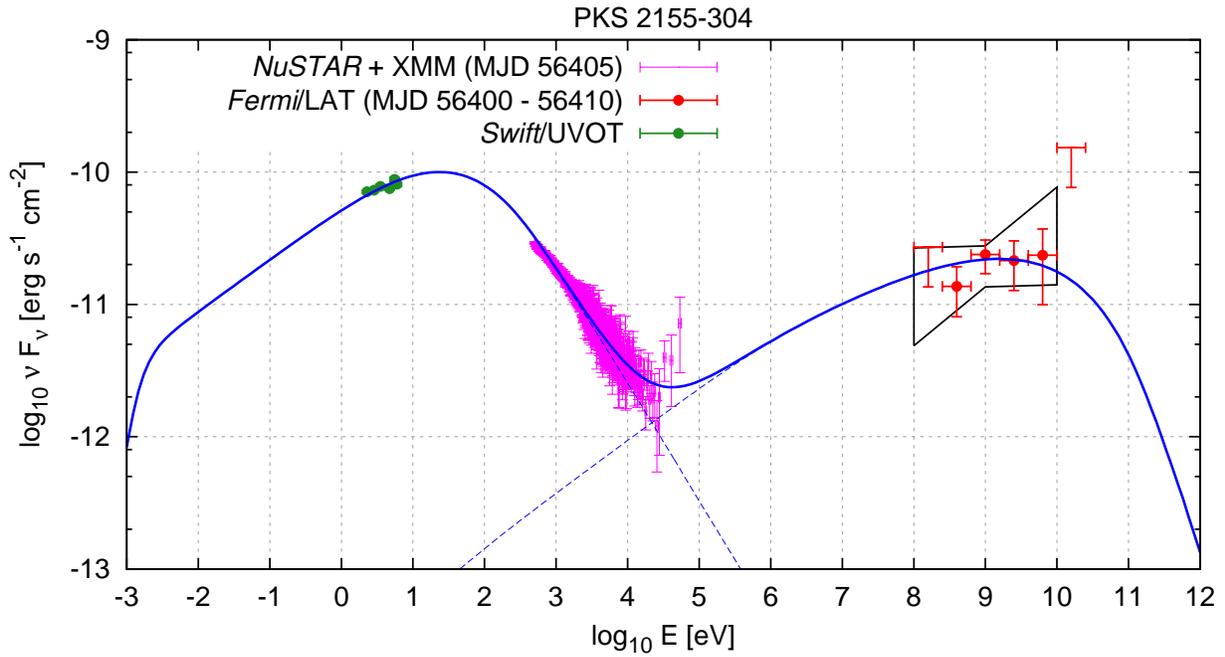}
\caption{Spectral energy distribution of PKS~2155-304 including contemporaneous data 
from {\it Fermi}-LAT, obtained for 10 days centered on the {\it NuSTAR} observation 
(with the data points in red, and a broad-band spectral fit ``butterfly'' in black), 
{\it NuSTAR}, XMM-{\it Newton} {\tt pn} (magenta). Also shown is a basic SED 
model including synchrotron and SSC components (blue).  We also plot the Swift-UVOT 
fluxes, corrected for reddening in our galaxy (green).}  
\label{fig4}
\end{figure}

During the {\it NuSTAR} observation on MJD~56405, PKS~2155-304 displayed a relatively 
low $\gamma$-ray state.  As mentioned above, we considered the data collected 
over 10-day period MJD~56400-56410.  A binned $E > 100$ MeV $\gamma$-ray spectrum, extracted 
from the {\it Fermi}-LAT data as described above, was fitted to a simple power-law model;  
{\it Fermi}-LAT measured the photon flux above $100\;{\rm MeV}$ to be 
$(8\pm 2)\times 10^{-8}\;{\rm ph\,s^{-1}\,cm^{-2}}$ with the photon index of $1.71\pm 0.15$.  
We plot the resulting data points as well as the fitted spectrum collected 
over the 10-day period in Figure \ref{fig4}.  


\section{Discussion:  modelling the broad-band spectral energy distribution and particle content of the jet}

In the context of the SSC models commonly invoked to explain 
the broad-band spectra and variability of HBL-type blazars, radio-through-soft 
X-ray emission is commonly attributed to the synchrotron process, with X-rays being 
due to the most energetic electrons.  The $\gamma$-ray emission is presumably produced 
via an inverse Compton process;  the commonly observed correlated variability in the 
VHE $\gamma$-ray and soft X-ray bands argues for a common energy range of the radiating 
particles.  Most commonly invoked models to describe the broad-band spectral energy 
distributions locate the cross-over between the synchrotron and inverse Compton peaks in 
the hard X-ray range, with the onset of the inverse-Compton peak manifesting itself as 
spectral hardening with increasing energy in the hard X-ray band.  Our {\it NuSTAR} spectrum 
of PKS~2155-304 provides evidence for such spectral hardening, which we interpret as 
the low-energy tail of the inverse-Compton component.  

We modelled the broad-band SED of PKS~2155-304 using 
the {\tt BLAZAR} code \citep{Mod03}.  Here, 
the X-ray spectrum is interpreted as the high-energy end of 
the synchrotron component, and the $\gamma$-ray spectrum is interpreted as the inverse-Compton 
part of the SSC component.  
Our main goal is to verify a hypothesis that the spectral hardening 
seen in the hard X-ray band could be due to the confluence between the synchrotron 
and inverse-Compton components.
Adopting a broken power-law distribution of injected electrons 
(${\rm d}N/{\rm d}\gamma \propto \gamma^{-p}$ with $p = p_2$ for 
$\gamma_{\rm min} < \gamma < \gamma_{\rm br2}$, and $p = p_3$ for $\gamma_{\rm br2} < \gamma < \gamma_{\rm max}$),
this hypothesis allows us to robustly constrain the low-energy index $p_2$ of 
the electron energy distribution, and the detailed X-ray spectrum allows us to constrain
the break Lorentz factor $\gamma_{\rm br2}$ and the high-energy index $p_3$ of the distribution.
In our basic model, we adopted the following parameters: jet Lorentz factor $\Gamma_{\rm j} = 15$,
magnetic field strength $B' = 0.5\;{\rm G}$ at a distance scale of $r = 0.065\;{\rm pc}$, 
jet half-opening angle and viewing angle $\theta_{\rm j} = \theta_{\rm obs} = 1/\Gamma_{\rm j}$,
and the emitting region radius $R = 1.3\times 10^{16}\;{\rm cm}$.
These parameters are consistent with those inferred from previous X-ray variability 
studies of this source (see, e.g., \citealt{Kat00};  \citealt{Fos07}; \citealt{Kat08};  
\citealt{Aha09}), but we also explored other values for $\Gamma_{\rm j}$, $B'$ and $r$ 
in order to verify that our key results do not depend on them (see below).  
The distribution of Lorentz factors of the injected electrons motivated by our hypothesis is 
characterized by $p_2 = 2.2$, $p_3 = 3.8$, $\gamma_{\rm min} \sim 1$, $\gamma_{\rm br2} = 2.6\times 10^4$ 
and $\gamma_{\rm max} \sim 10^7$.
The very high value of $\gamma_{\rm max}$ is not constrained by any observational data.
With this power-law index $p_2$,
most of the electron power is contained in the lowest-energy electrons,
and hence the average Lorentz factor of the injected electrons is only 
$\left<\gamma_e\right> \simeq 5.6$.
This model is presented in Figure \ref{fig4},
and it predicts the synchrotron self-absorption break 
at $\sim 0.7\,{\rm mm}$ (or $1.77 \times 10^{-3}\,{\rm eV}$).    

We calculate components of the jet power required by the model as 
$L_q = \pi R^2\Gamma_{\rm j}^2u_q'c$,
where $u_q'$ is the energy density of quantity $q$ measured in the jet comoving frame.
In particular, we find magnetic power $L_B \simeq 3.8\times 10^{43}\;{\rm erg\,s^{-1}}$,
electron power $L_e \simeq 3.7\times 10^{44}\;{\rm erg\,s^{-1}}$,
and radiation power $L_r \simeq 1.7\times 10^{43}\;{\rm erg\,s^{-1}}$.
These values suggest that the jet composition is strongly dominated by matter with 
$L_e / L_B \simeq 10$, without taking into account any protons.
However, one must consider charge neutrality, so each electron 
must have a corresponding proton or positron.  
Assuming -- for now -- one cold proton per electron, this would predict a very high proton power
of $L_p \sim (m_p/m_e)(L_e/\left<\gamma_e\right>) \sim 1.2\times 10^{47}\;{\rm erg\,s^{-1}}$.  
Providing such a large amount of kinetic power via accretion is unrealistic for an HBL-type blazar.
Even if the mass of the black hole  M$_{\rm BH}$ is $10^9$ solar masses (not directly measured, but 
adopted by, e.g., \citealt{Aha07}), and assuming high efficiency of conversion of the 
accretion power to jet power, in order to provide 
the kinetic power for the jet with equal number of electrons and protons,
the accretion rate would have to exceed the Eddington rate.
Such a high accretion rate would result in an optically thick accretion disk, 
which in turn should reveal quasi-thermal components (emission lines, blue bump, 
and possible Compton reflection component), as commonly seen in high accretion 
rate active galaxies.  This is in conflict with the absence of such thermal components in
HBL-type blazars (and in PKS~2155-304 in particular), which in turn 
suggests that HBL-type blazars accrete via inefficient, low accretion-rate, 
advection-dominated flows, or ADAFs (for a recent overview, see \citealt{Yuan14}).

Alternatively, we can assume the presence of electron-positron pairs and estimate
the effective numbers of leptons per proton $n_{\rm e+e-}/n_{\rm p}$.
Since in any case the jet 
appears to be matter-dominated, we assume that the power of leptons $L_{\rm e+e-}$ originates 
from the dissipation of the power of protons $L_p$, such that $L_{\rm e+e-} = \eta_{\rm diss,e} \, L_p$, 
where $\eta_{\rm diss,e} \sim 0.1$ is the fraction of the dissipation efficiency measuring 
the energy fraction transferred to the leptons.
Therefore, we find 
$n_{\rm e+e-}/n_{\rm p} \simeq (m_p/m_e)(\eta_{\rm diss,e}/\left<\gamma_e\right>) \sim 33$ 
and $L_p \sim 4\times 10^{45}\;{\rm erg\,s^{-1}}$.
We note that a similar constraint -- but using different arguments -- was obtained 
for the pair content in luminous blazars associated with flat spectrum radio quasars 
(FSRQs) by \cite{Sik00}, although their conclusions were recently somewhat weakened \citep{Sik13}.
The problem of pair content in blazar jets was also investigated by \cite{Ghi12}, 
and most recently, \cite{Ghi14} concluded that the presence of a significant number 
of pairs in the powerful blazars is unlikely.  In any case, FSRQ jets are unlikely 
to be proton-free, as jets consisting of pure pairs would overproduce the observed 
X-ray flux, via bulk-Compton scattering of ambient, circum-nuclear photons.  
Since in HBL-type blazars such ambient photon fields are very weak or absent,
the {\sl minimum} proton content is basically unconstrained.
We also note that, since the relative number of pairs $n_{\rm e+e-}/n_{\rm p}$ depends 
only on the dissipation efficiency $\eta_{\rm diss,e}$ and on the average electron energy 
$\left<\gamma_e\right>$, our key result depends primarily on our assumptions on the 
energy distribution of injected pairs.  Still, in order to determine the sensitivity 
of the resulting $\left<\gamma_e\right>$ (which in turn
determines $n_{\rm e+e-}/n_{\rm p}$) on our adopted $\Gamma_{\rm j}$ and $r$,
we attempted two additional models.  In one case, we adopted $\Gamma_{\rm j} = 20$,
and in another, $r = 0.13$ pc;  in both cases, we kept other parameters
as above, but we adjusted $B'$ and $\gamma_{\rm br2}$ to make the model 
agree with the data.  The resulting $\left<\gamma_e\right>$ varied by respectively (roughly) 
$-2 \%$ and $+0.5 \%$ (with correspondingly small changes 
in $n_{\rm e+e-}/n_{\rm p}$), certainly not sufficient to bring this ratio close to unity.  Thus, 
by imposing charge neutrality, a jet consisting of pure electron-proton plasma, with no positrons, 
is not favored.  

Alternatively, we can consider a reduced number of the low-energy electrons, 
which is basically unconstrained by the observational data.
If the electrons responsible for the hard X-ray part of the SSC component 
($\gamma \sim 300$) were in the fast-cooling regime, we could postulate a 
significantly harder low-energy power-law index $p_2 \simeq 1.2$ for the 
injected electron population.  However, in our model the electrons are cooling 
efficiently only for $\gamma > 10^4$.  Therefore, in order to reduce the number 
of low-energy electrons, it is necessary to introduce a second break in the 
electron injection spectrum with $p = p_1$ for $\gamma_{\rm min} < \gamma < \gamma_{\rm br1}$, 
and $p = p_2$ for $\gamma_{\rm br1} < \gamma < \gamma_{\rm br2}$.  For example, adopting 
$p_1 = 1$ and $\gamma_{\rm br1} = 100$, we have $\left<\gamma_e\right> \simeq 84$ 
and $L_{\rm e+e-} \simeq 1.5\times 10^{44}\;{\rm erg\,s^{-1}}$, and hence 
$n_{\rm e+e-}/n_{\rm p} \sim 2.2$ and $L_p \sim 1.5\times 10^{45}\;{\rm erg\,s^{-1}}$.
Other possibilities, e.g., a sharp low-energy cutoff with $\gamma_{\rm min} \sim 100$, can also be considered.
In either of the above cases, assuming the presence of either pairs or a low-energy break
can bring the required jet power to reasonable values.
However, without changing other parameters such as $\Gamma_{\rm j}$,
it is very challenging to bring the jet composition closer to equipartition, 
as these parameters do not affect the magnetic power $L_B$.

\section{Conclusions}

PKS~2155-304 displayed a relatively low state during the first {\it NuSTAR} observations 
of the source in April 2013, with the measured 2 -- 10 keV X-ray flux of only 
$\sim 1.1 \times 10^{-11}$ erg cm$^{-2}$ s$^{-1}$, roughly three times 
lower than the lowest X-ray flux in August-September 2008, reported by \cite{Aha09}.  
{\it NuSTAR} data reveal a steep ($\Gamma \sim 3$) spectrum below $\sim 10$ keV, hardening to 
$\Gamma \sim 2$ above $\sim 10$ keV.  When fitted with strictly simultaneous XMM-{\it Newton} data, 
the soft component is best-fit as a log-parabolic model, and the hard tail is
even more significant.  
It is naturally expected that such spectral hardening as we detect in the combined 
{\it NuSTAR} and XMM-{\it Newton} data would be more easily detectable when the source is in a state 
of a relatively low soft X-ray flux.  This is because the soft X-ray and VHE $\gamma$-ray 
variability in HBL BL Lacs is generally more rapid and has larger amplitude than 
that at lower energy of the respective peaks.  This is partially due to more rapid energy losses 
with increasing particle energy.  Therefore, the chance of detecting the presumably 
less variable onset (low-energy end) of the inverse Compton component is actually {\sl greater} when 
the high-energy tail of the synchrotron peak is weak, and does not dilute the Compton component.  
Indeed, our data taken in an extremely low-flux state reveal such a component.    

An application of the SSC model allows us to estimate the particle content in the jet.  
If we assume one proton per electron, then the total power of the jet is dominated by 
two orders of magnitude by particles, amounting to $L_p \sim 10^{47}\;{\rm erg\,s^{-1}}$.  
This would require a very large amount of power to be delivered via accretion,
and would imply accretion at a highly super-Eddington rate.  This, in turn, is unlikely given the  
absence of any quasi-thermal spectral components one would expect to be present 
in the optical/UV spectra of this source.  Therefore, we consider a more plausible scenario,
where the jet contains significantly more than one lepton per proton, meaning that 
by number, the jet is dominated by electron-positron pairs.  This allows 
the reduction of the required jet power by two orders of magnitude, bringing it 
to more realistic values.  The required jet power can also be reduced by introducing 
an additional break in the electron injection spectrum, e.g., with $\gamma_{\rm br1} \sim 100$ 
and $p_1 = 1$.  In either case explored here, the total power of the jet is 
dominated by particles rather than by magnetic fields.  

In summary, while the presence of electron-positron pairs was 
previously postulated in relativistic jets of FSRQs 
(see \citealt{Sik00}), the new constraint from {\it NuSTAR} on the low-energy 
part of the electron distribution suggests that copious pairs may be present 
in jets associated with the lineless, HBL-type blazars.

\acknowledgements
The {\it Fermi}-LAT Collaboration acknowledges support for LAT development, 
operation and data analysis from NASA and DOE (United States), CEA/Irfu and 
IN2P3/CNRS (France), ASI and INFN (Italy), MEXT, KEK, and JAXA (Japan), and 
the K.A.~Wallenberg Foundation, the Swedish Research Council and the National 
Space Board (Sweden). Science analysis support in the operations phase from 
INAF (Italy) and CNES (France) is also gratefully acknowledged.  This work was 
partially supported under the NASA contract no.\ NNG08FD60C, and made use of 
observations from the NuSTAR mission, a project led by California Institute 
of Technology, managed by the Jet Propulsion Laboratory, and funded by NASA.  
We thank the {\it NuSTAR} Operations, Software and Calibration teams for support 
of the execution and analysis of these observations. This research has made 
use of the {\it NuSTAR} Data Analysis Software (NuSTARDAS) jointly developed by 
the ASI Science Data Center (ASDC, Italy) and the California Institute of Technology 
(USA).  K.N. was supported by NASA through Einstein Postdoctoral Fellowship 
grant number PF3-140130 awarded by the Chandra X-ray Center, and by the Polish 
National Science Centre grant 2015/18/E/ST9/00580.
M.\,B. acknowledges support from NASA under the NASA Earth and 
Space Science Fellowship Program, grant NNX14AQ07H.


\clearpage

\end{document}